\documentclass[10pt]{iopart}

\usepackage{graphicx}
\usepackage{hyperref}
\usepackage{xcolor}
\usepackage{upgreek}

\begin{document}

\title[Voigt transmission windows in optically thick atomic vapours]{Voigt transmission windows in optically thick atomic vapours: a method to create single-peaked line centre filters}

\author{Jack~D~Briscoe, Fraser~D~Logue, Danielle~Pizzey, Steven~A~Wrathmall and Ifan~G~Hughes}

\address{\textit{Physics Department, Durham University, South Road, Durham, DH1 3LE, United Kingdom.}}

\ead{jack.d.briscoe@durham.ac.uk}

\begin{abstract} 
Cascading light through two thermal vapour cells has been shown to improve the performance of atomic filters that aim to maximise peak transmission over a minimised bandpass window. In this paper, we explore the atomic physics responsible for the operation of the second cell, which is situated in a transverse (Voigt) magnetic field and opens a narrow transmission window in an optically thick atomic vapour. By assuming transitions with Gaussian line shapes and magnetic fields sufficiently large to access the hyperfine Paschen-Back regime, the window is modelled by resolving the two transitions closest to line centre. We discuss the validity of this model and perform an experiment which demonstrates the evolution of a naturally abundant Rb transmission window as a function of magnetic field. The model results in a significant reduction in two-cell parameter space, which we use to find theoretical optimised cascaded line centre filters for Na, K, Rb and Cs across both D lines. With the exception of Cs, these all have a better figure of merit than comparable single cell filters in literature. Most noteworthy is a Rb-D2 filter which outputs $>92\%$ of light through a single peak at line centre, with maximum transmission $0.71$ and a width of 330~\rm{MHz} at half maximum. 
\end{abstract}

\noindent{\it Keywords\/}: Magneto-optical effects, atomic filter, atomic spectroscopy, hyperfine Paschen-Back, atomic transitions, optimisation

\ioptwocol

\section{Introduction}
Manipulation of light using magneto-optical effects is used frequently in quantum optics literature~\cite{caltzidis2021atomic, higgins2021electromagnetically, siverns2019demonstration, budker2002resonant, auzinsh2010optically}. Popular applications include optical isolators~\cite{weller2012optical, aplet1964faraday}, magnetometry~\cite{budker2007optical, sutter2020recording, kitching2018chip, auzinsh2022wide} and laser frequency stabilisation~\cite{nagourney2014quantum}. One of the main applications is narrow-band magneto-optical filters which only permit light in the spectral vicinity of atomic resonances~\cite{gerhardt2018anomalous}. These atomic filters utilise atom-light interactions in a variety of configurations, with the two most common being Faraday~\cite{dick1991ultrahigh, menders1991ultranarrow, yin1991theoretical, chen1993sodium, Harrell:09, yan2022dual} and Voigt filters~\cite{menders1992blue, kudenov2020dual, yin2016tunable}. The names are inherited from the corresponding Faraday and Voigt effects, where light is magneto-optically rotated due to induced birefringence as it traverses through a dispersive medium~\cite{faraday1846experimental}. While based upon the same underlying principles, these effects differ by the relative orientation of the light propagation vector $\vec{k}$ and magnetic field vector $\vec{B}$~\cite{ponciano2020absorption}. As a result, an atomic medium will exhibit different propagation eigenmodes, thus modifying the profiles produced by these filter geometries~\cite{keaveney2018self}. Single cell atomic filters can also be realised in unconstrained $\vec{B}$ geometries, where the angle that $\vec{B}$ makes with $\vec{k}$ can be between 0$^{\circ}$ (Faraday) and 90$^{\circ}$ (Voigt)~\cite{higgins2020atomic, keaveney2018optimized}. However, the wave equation solutions are significantly more complex~\cite{keaveney2018elecsus, rotondaro2015generalized}, and consequently these filters are rarely explored.

Atomic filters are employed in an ever-growing range of applications, including single photon filtering~\cite{siyushev2014molecular, portalupi2016simultaneous}, atmospheric lidar~\cite{yong2011flat, li2012doppler}, designing frequency-selective lasers~\cite{shi2020dual, chang2022frequency, tang202118w, keaveney2016single}, ghost imaging~\cite{yin2022using}, and optical communication~\cite{junxiong1995experimental, zhang2021background}. In solar physics studies, filters are realised by cascading light through multiple thermal vapour cells~\cite{erdelyi2022solar,cacciani1978magneto, cacciani1990solar}. Recently, we have shown cascading can also be used to significantly improve the performance of atomic filters~\cite{logue2022better} that aim to maximise the figure of merit (FOM)~\cite{kiefer2014faraday}
\begin{equation}
\rm{FOM} = \frac{\mathcal{T}(\nu_{\rm{max}})}{\rm{ENBW}}~;~\rm{ENBW} = \frac{\int\mathcal{T}(\nu)\rm{d}\nu}{\mathcal{T}(\nu_{\rm{max}})}~,
\end{equation}
where $\mathcal{T}$ is the transmission relative to the input light intensity. We use FOM as it balances maximising peak transmission $\mathcal{T}(\nu_{\rm{max}})$, while simultaneously minimising equivalent noise bandwidth (ENBW) over a large frequency ($\nu$) range. Therefore, we actively search for filters which are optically thick at frequencies detuned away from peak transmission. The key assumption we make is that the two cells in a cascaded system have independent roles: the role of the first cell is to magneto-optically rotate light and create regions of high peak transmissions, while the role of the second cell is to absorb light away from this peak, thus reducing ENBW. We therefore lift the fundamental limit of single cell filters, which need to both rotate and absorb light using the same cell parameters~\cite{gerhardt2018anomalous}.

In this paper, we show that transmission windows in optically thick atomic vapours can be used to create single-peaked line centre filters, by acting as the ideal second cell in a cascaded cell atomic filter. In contrast to our previous experimental realisation with Rb~\cite{logue2022better}, the focus of this paper is to use the atomic physics and thermal properties of optically thick vapours, situated in the Voigt geometry~\cite{ponciano2020absorption, keaveney2019quantitative}, in order to generalise our complex cascaded system to other alkali metals and atomic transitions. By resolving thermally broadened atomic transitions~\cite{siddons2009off, weller2011absolute} in the hyperfine Paschen-Back (HPB) regime~\cite{sargsyan2022saturated, sargsyan2012hyperfine, olsen2011optical, zentile2014hyperfine, weller2012absolute}, a simple two transition model is derived which describes the mechanism to open the window. We discuss the validity of this model, which assumes transitions with Gaussian line shapes, and use experiment to visualise why it breaks down. The same experiment demonstrates the typical evolution of a Voigt transmission window as a function of magnetic field strength. Finally, the model is used to significantly reduce two-cell parameter space, and therefore we find theoretical optimised cascaded line centre filters for Na, K, Rb and Cs. With the exception of Cs, these exceed the performance of single cell atomic filters in literature, as measured by FOM~\cite{gerhardt2018anomalous, keaveney2018optimized, zentile2015atomic}\footnote{Better FOMs can be achieved using cold atoms and velocity selection~\cite{guancold}. The cost is significantly reduced peak transmissions, and setups that rely on pump lasers, magneto-optical traps and additional control systems~\cite{guancold, zhuang2021ultranarrow, bi2016ultra}.}. It is clear that the second cell absorption profiles of cascaded filters result in significant reductions in noise displaced away from line centre. We therefore expect there will be great utility in applying transmission windows to single cell atomic filters at larger optical powers, which are influenced by signal intensity away from line centre~\cite{luo2018signal}.

The paper is structured as follows: Section~\ref{Sec: Theory} discusses the relevant theory for analysing optically thick atomic vapours in the HPB regime; operation of the cascaded cell line centre filter is shown in Section~\ref{Sec: TwoCells}; the two transition model for optically thick transmission windows is presented in Section~\ref{Sec: Window}, supported by experiment. The paper culminates in Section~\ref{Sec: OptimizedFilters}, where we show optimised cascaded cell filters which output the majority of light through  a single peak at line centre. After comparison to literature, the paper concludes in Section 6.

\section{Theoretical model} 
\label{Sec: Theory}
To study optically thick atomic vapours, we use $ElecSus$~\cite{zentile2015elecsus, keaveney2018elecsus}. $ElecSus$ is an open-source computer program that calculates the electric susceptibility $\chi$ of alkali metal atoms in a thermal vapour, assuming closed two level systems and a weak probe regime where population transfer is minimised~\cite{sherlock2009weak}. Absorption by the atoms is calculated using the imaginary part of $\chi$. This is due to the electric field extinction term $\rm{exp}$(-$\chi_{\rm{I}}kz/2$) which attenuates light as it propagates a distance $z$ through a medium~\cite{etheses7747}. Comparing this term with the Beer-Lambert law gives the absorption coefficient
\begin{equation}
\alpha_{j}(B,T,\Delta) =  k\frac{C_{j}^{2}(B)d^{2}}{2(2I + 1)} \frac{\mathcal{N}(T)}{\hbar\epsilon_{0}}\mathcal{V}(\Delta, T)~,
\label{eqn:alpha}
\end{equation}
which gives the line shapes of profiles detuned $\Delta$ from each resonance $j$ in an atomic system~\cite{Siddons_2008}. Each Voigt profile $\mathcal{V}(\Delta, T)$ is scaled vertically by temperature ($T$) due to the exponential behaviour of alkali-metal number densities $\mathcal{N}$~\cite{wu1986optical}, while simultaneously being thermally broadened along the horizontal detuning axis by Doppler motion~\cite{preston1996doppler} and self-broadening mechanisms~\cite{weller2011absolute, lewis1980collisional}. Profiles are also scaled vertically by the wavevector magnitude $k$, nuclear spin $I$, and the non-trivial behaviour of line-strength factors $C_{j}^{2}d^{2}$, where $C_{j}^{2}$ are a function of magnetic field strength~($B$)~\cite{etheses7747}. The Voigt profile is a convolution of Gaussian and Lorentzian functions~\cite{tudor1963new}, and in certain regimes can take one of these line shapes by approximation. The Lorentzian approximation is valid far-off resonance, whereas the Gaussian approximation is valid for $\Delta < 1.5~\omega_{\rm{D}}$~\cite{siddons2009off}. We show our model satisfies this constraint in Section~\ref{Subsec: Model}, and therefore each Voigt profile is approximated by~\cite{siddons2009off}
\begin{equation}
\mathcal{V}(\Delta, T) \sim \frac{2\sqrt{\pi \ln{2}}}{\omega_{\rm{D}}}\rm{exp}[-(\Delta/\omega_{D})^{2}4\ln{2}]~,
\label{eqn:Gaussian}
\end{equation}
where $\omega_{\rm{D}} = 2\sqrt{\ln{2}}\,ku$ is the FWHM Doppler width of each Gaussian profile, and $u \propto \sqrt{T}$ is the FWHM of the velocity distribution describing the thermal motion of the atoms~\cite{foot2004atomic}.

While $T$ and $B$ can modify the shape of an atomic spectrum, its position intrinsically depends on $B$ due to the evolution of transition energies via the Zeeman effect~\cite{woodgate1970elementary}. By using a matrix representation to construct the Hamiltonian $\hat{H}$ describing our atomic system, we are able to model the Zeeman interaction, as well as the effects of fine and hyperfine structure. We calculate $\hat{H}$ in a completely uncoupled basis, and then diagonalise $\hat{H}$ to give its eigenenergies and eigenstates. Transition energies are calculated as the difference between excited and ground state eigenenergies for all electric dipole-allowed transitions between uncoupled basis states. Therefore, the positioning of an atomic spectrum depends entirely upon the quantum numbers, atomic constants and magnetic field that contribute towards $\hat{H}$. For an in-depth analysis of the matrix representation used by $ElecSus$, see~\cite{etheses7747}. The theory for how $\chi$ relates to a filter's transmission profile is discussed in literature, for example~\cite{dressler1996theory}.

The behaviour of states can be split into regimes defined by the magnitude of $B$~\cite{weller2012absolute}. At large magnetic fields, we enter the HPB regime where energy levels are described by uncoupled basis states $|I, m_{I}, J, m_{J}\rangle$. In this notation, $J$ describes the total electronic spin, $m$ are projection quantum numbers, and for each state labelled $m_{J}$ there are $2I + 1$ energy levels~\cite{zentile2014hyperfine, weller2012absolute}. We assume the transmission window opens in the HPB regime, and therefore use a strong $B$ field approximation to determine state eigenenergies~\cite{woodgate1970elementary}
\begin{equation}
E_{\vert I, m_{I}, J, m_{J} \rangle} = A\,m_{I}m_{J} +\mu_{\rm{B}}B(g_{I}m_{I} + g_{J}m_{J})~.
\label{eqn: strongEnergy}
\end{equation}
The energy $E$ of the state $\vert I, m_{I}, J, m_{J} \rangle$ is in units of $h$, $A$ is the magnetic dipole constant, $\mu_{\rm{B}}$ is the Bohr magneton and the $g$-factors are gyromagnetic ratios. Since $g_{I} << g_{J}$, the effect of an external magnetic field in the HPB regime is to split the state eigenenergies into groups defined by $m_{J}$.

\begin{figure*}[t]
\centering
{\includegraphics[width=\linewidth]{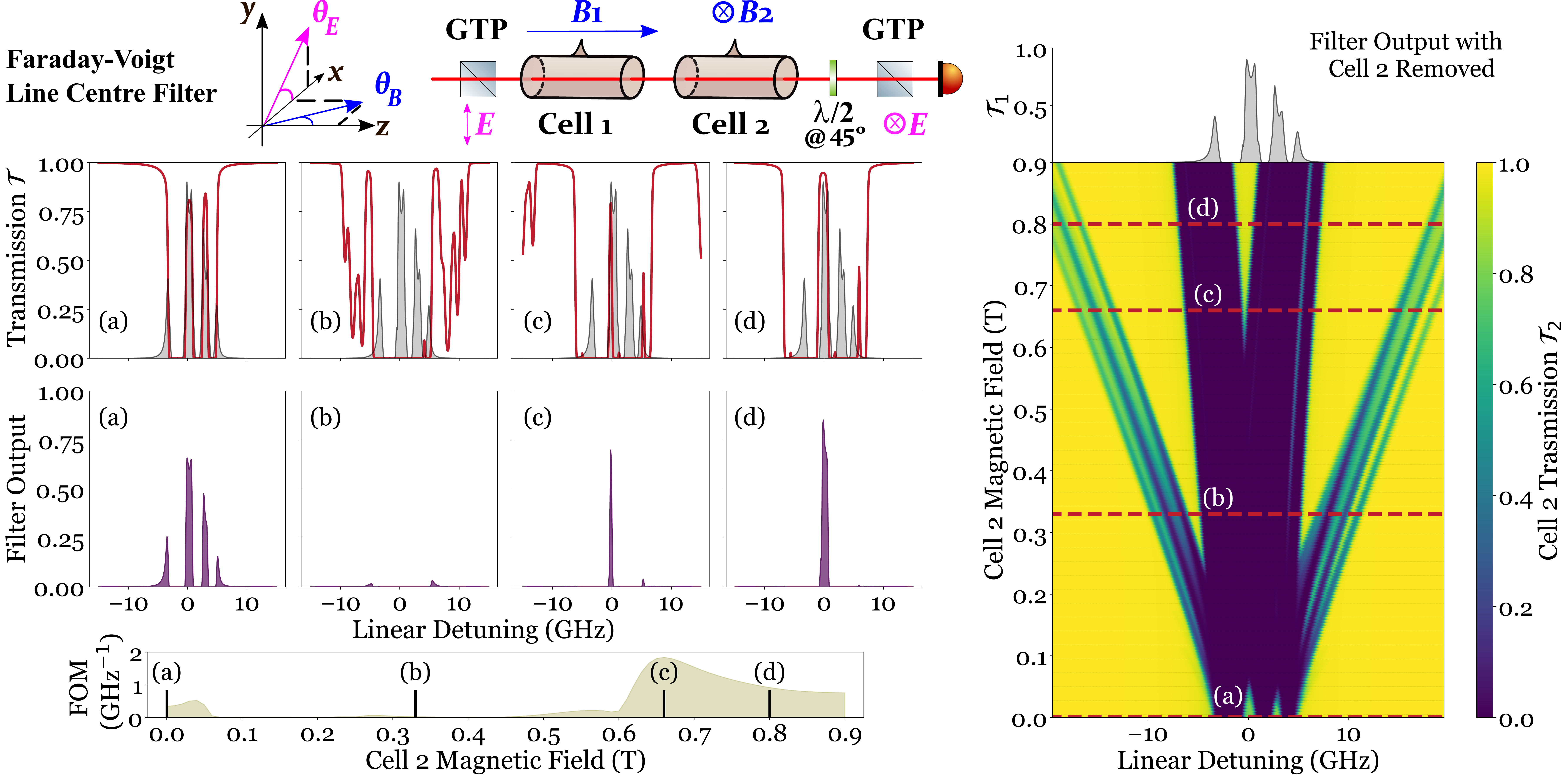}}
\caption{A Faraday-Voigt line centre filter is realised by cascading light through thermal vapour cells in the Faraday ($\theta_{B} = 0^{\circ}$) and Voigt ($\theta_{B} = 90^{\circ}$) geometries, with both cells placed between crossed polarisers (GTP). We show the evolution of a natural abundance Rb-D2 Faraday-Voigt filter, depicted at second cell magnetic fields ($B_{2}$) of (a) $0~\rm{T}$, (b) $0.33~\rm{T}$, (c) $0.66~\rm{T}$, and (d) $0.8~\rm{T}$. Each filter is displayed in a column and has fixed parameters: $T_{1} = 105~^{\circ}\rm{C}$, $B_{1} = 8.6~\rm{mT}$, $T_{2} = 130~^{\circ}\rm{C}$, and cell lengths $5~\rm{mm}$. Vertical linearly polarised light is input through the first cell ($\theta_{E} = 90^{\circ}$). In each column, we display filter output (purple) in the bottom plot, while the top plot shows both the second cell spectra (red) and filter output if the second cell is removed (grey). A transmission heatmap shows the full evolution of the second cell spectra with increasing $B_{2}$. In particular, it shows the second cell spectrum splitting at approximately $0.6~\rm{T}$, highlighted by a sharp rise in FOM (olive) which peaks at $1.84~\rm{GHz}^{-1}$. The theoretical model has been experimentally verified~\cite{logue2022better}.}
\label{fig: RbLineCentre}
\end{figure*}

\section{Faraday-Voigt line centre filter} \label{Sec: TwoCells}

The setup for our cascaded line centre filter has two thermal vapour cells placed between crossed polarisers, as is shown in figure~\ref{fig: RbLineCentre}. The first cell is placed in the Faraday geometry ($\vec{k}\parallel\vec{B}$, $\theta_{B} = 0^{\circ}$), which then outputs light to a second cell in the Voigt geometry ($\vec{k}\perp\vec{B}$, $\theta_{B} = 90^{\circ}$). We use the Voigt geometry since each of the two refractive index solutions to the wave equation are associated with a propagation eigenmode that can independently couple to $\sigma^{+/-}$ transitions, $\pi$ transitions, or all three simultaneously~\cite{ponciano2020absorption, keaveney2018self}. The relative strength of these transitions depends on the orientation of the atom-light system. In the Voigt geometry, we define the angle of input polarisation $\theta_{E}$ relative to $\vec{B}$ (see figure~\ref{fig: RbLineCentre}). Therefore, $\pi$ transitions are driven with a strength weighted by $\rm{cos}^{2}$$\theta_{E}$, and $\sigma^{+/-}$ transitions are driven with a strength weighted by $\rm{sin}^{2}$$\theta_{E}$~\cite{ponciano2020absorption}. For large enough optical depths and $\theta_{E} \sim 0^{\circ} (90^{\circ}$), it is therefore possible for atoms exhibiting $\pi$ ($\sigma^{+/-}$) transitions to absorb 100\% of light.  This is important as the aim of the second cell is to create an absorption profile which is optically thick at the largest possible frequency range detuned away from the filter's peak transmission. This contrasts with the Faraday geometry at large $B$, where only 50\% of light  can be absorbed on resonance due to the dichroism associated with each propagation eigenmode~\cite{caltzidis2021atomic}.

In figure~\ref{fig: RbLineCentre}, we show plots for a theoretical natural abundance Rb Faraday-Voigt filter, analogous to the one constructed in \cite{logue2022better}. Any number following a parameter refers to the cell i.e Faraday $= 1$, Voigt $= 2$. In this example, we input vertical linear light into the first cell (Faraday, $L_{1} = 5~\rm{mm}$, $\theta_{E} = 90^{\circ}$), which by our independent cell assumption rotates the light by $90^{\circ}$ such that horizontal linear light passes through the second cell (Voigt, $L_{2} = 5~\rm{mm}$). Since $\vec{E} \parallel \vec{B}$ through the second cell, its spectrum exhibits $\pi$ transitions only, whereas we select for both $\sigma^{+}$ and $\sigma^{-}$ transitions in~\cite{logue2022better}. At zero magnetic field (a), the second cell has a standard Rb-D2 spectrum (red). Typically, we would see 4 groups of resolved transition profiles~\cite{Siddons_2008}, but due to the high second cell temperature $T_{2} = 130^{\circ}\rm{C}$, the profiles of the two leftmost groups of transitions merge. In column (b), where $B_{2} = 0.33~\rm{T}$, the filter has no output (purple) since the optically thick second cell spectrum is covering the first cell transmission (grey). As $B_{2}$ increases, weaker transitions begin to resolve from the spectrum, decreasing in amplitude due to diminishing $C_{j}^{2}$. In column (c), the optimised filter is realised at $B_{2} = 0.66~\rm{T}$ as the second cell absorption profile begins to split at approximately $0.60~\rm{T}$. This is highlighted in the spectrum heatmap, displayed on the right side of the figure. At this critical magnetic field, the transmission window opens as the two main groups of strong Rb-D2 $\pi$ transitions separate due to different transition line gradients (see Section~\ref{Subsec: Model}). Column (d) highlights the sensitivity of this type of filter. A larger value of $B_{2}$ opens the transmission window further: while this allows for a filter with a larger peak transmission, the FOM decreases (olive) due to larger increase in ENBW.  The output is an ultra-narrow single-peaked line centre filter which forms in an optically thick Voigt transmission window.

\section{Opening the transmission window} \label{Sec: Window}
\subsection{Two transition model} \label{Subsec: Model}
A large magnetic field is required to create the split in an optically thick second cell absorption profile, opening the window for first cell transmission at line centre. The split occurs because the energies of an atom's ground and excited hyperfine states separate into groups defined by their $m_{J}$ quantum numbers for strong $B$ fields, in accordance with equation~\ref{eqn: strongEnergy}. The behaviour of both the ground and excited state energies imprint onto the transition energies for all electric-dipole transitions of an atomic system, and therefore splits emerge between different groups of transitions. This is described in \ref{App1}.

\begin{figure}[t]
\centering
{\includegraphics[width=0.9\linewidth]{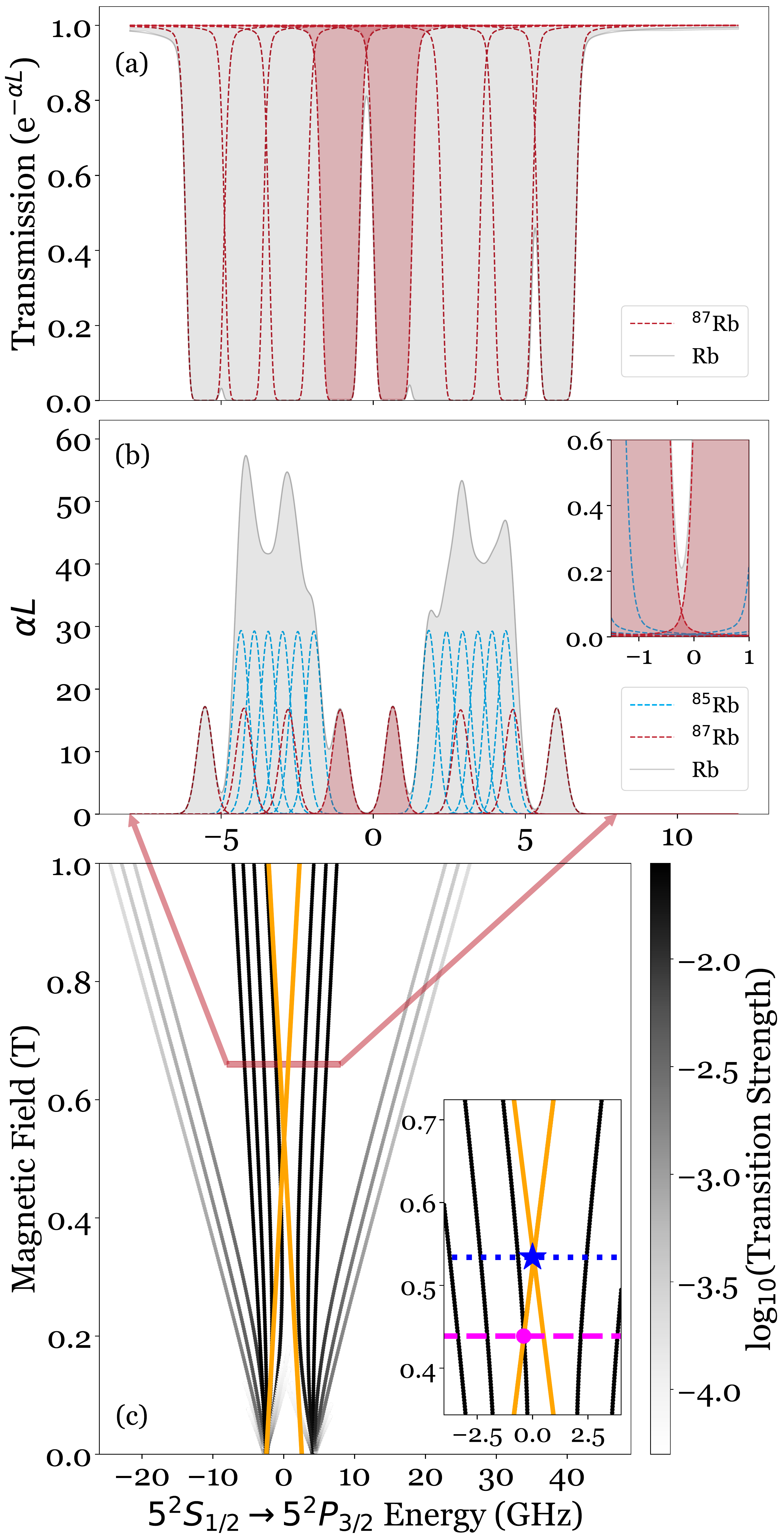}}
\caption{(c): Evolution of $^{87}$Rb $\pi$ transition energies with increasing magnetic field, coloured by the logarithmic transition strength. The separation of strong transitions defined by $m_{J} = \pm1/2 \rightarrow \pm1/2$ opens a transmission window in optically thick $^{87}$Rb and Rb spectroscopy. Also plotted are the energies of the inner transitions using a strong $B$ field approximation (orange), defined by $m_{I} = -I$, and their crossing point is shown in the inset (blue, dotted). The exact crossing point of these transitions occurs at a smaller $B$ (magenta, dashed). (b): Optical depth $\alpha L$ for natural abundance Rb, $T = 130^{\circ}$C, $B = 0.66~\rm{T}$, $L = 5~\rm{mm}$, in a geometry where $\pi$ transitions are induced. The contributions to the total profile (grey) from each isotope are identified by colour ($^{85}$Rb blue, $^{87}$Rb red). (a): The transmission spectrum $\mathcal{T} = \rm{exp}$($-\alpha L$) of the same parameters. Only the $^{87}$Rb and total contributions are shown. In both (a) and (b), the two profiles closest to zero detuning are highlighted (red), as they are responsible for the evolution of the transmission window.}
\label{fig: D2sigmaTransitionsCrossing}
\end{figure}

There is an abundance of literature which investigates transition energies both theoretically and experimentally~\cite{tremblay1990absorption, umfer1992investigations, windholz1985zeeman, windholz1988zeeman}. The general application is precision magnetometry~\cite{budker2007optical, ponciano2020absorption}, where stretched states are typically used as atomic frequency references at very large magnetic fields~\cite{george2017pulsed, ciampini2017optical}.
In our system, we model the splitting point as the total magnetic field $B(T)$ across the Voigt cell required to resolve the two thermally broadened transitions closest to zero detuning. The first component of $B(T)$ is the initial magnetic field required for those transitions to cross, which depends only on quantum numbers and atomic constants i.e. atomic physics. We call this component the strong cross $B_{\rm{SC}}$ as it is found by assuming the HPB regime. In this section, we consider the example where atoms exhibit $\pi$ transitions on the D2 line, and plot these
for $^{87}$Rb in figure~\ref{fig: D2sigmaTransitionsCrossing}c. The colour of each transition energy indicates the logarithmic strength of the transition ($\propto C_{j}^{2}$), such that we can identify the two groups of strong transitions (darker lines) in the HPB regime. By using equation~\ref{eqn: strongEnergy} and the notation $m'_{J} = m_{J} + q$, we can derive equation~\ref{APDXeqn: strongEnergy3} which approximates transition energies in the HPB regime
\begin{eqnarray}\label{eqn: strongTransitionEnergy}
\Delta E(q) & = &  m_{I}[m_{J}(A' - A) + qA'] \\ \nonumber  
& + & \mu_{\rm{B}}B[m_{J}(g'_{J} - g_{J}) + qg'_{J}]~.
\end{eqnarray}
In equation~\ref{eqn: strongTransitionEnergy}, we have used primes to indicate the excited state, and $q = 0$ ($\pm 1$) for $\pi$ ($\sigma^{\pm}$) transitions. In \ref{App1}, we identify the $m_{J}$ quantum numbers responsible for each transition group, and isolate the $m_{I}$ values of the inner transitions closest to zero detuning. By substituting these into equation~\ref{eqn: strongTransitionEnergy} and equating the two transitions, we derive
\begin{equation}
B_{\rm{SC, D2}}^{\pi} = \frac{3}{2}\frac{(A - A')I}{\mu_{\rm{B}}}~,
\label{eqn: strongPiCrossD2}
\end{equation}
for the D2 line. All equations for $\pi$ and $\sigma^{\pm}$ transitions across both D lines can be found in \ref{App1} (equations~\ref{APDXeqn: strongPiCrossD2}-\ref{APDXeqn: strongCrossOther}). We add to figure~\ref{fig: D2sigmaTransitionsCrossing}c the strong $B$ energy approximation (orange) of the inner transitions using equation~\ref{eqn: strongTransitionEnergy}, the identified quantum numbers from \ref{App1} and atomic constants from~\cite{zentile2015elecsus}. The bottom inset shows $B_{\rm{SC, D2}}^{\pi}$ (blue, dotted), as well as the exact cross of the two transition groups determined analytically using $ElecSus$ (magenta, dashed). The ratio of the strong and exact crossing points is 1.217, which we find is constant for Na, $^{39}$K, and $^{87}$Rb ($I = 3/2$). As $I$ increases, this ratio tends to 1, and therefore can be viewed as a correction for using the HPB strong $B$ approximation.

The second component of $B(T)$ gives the additional magnetic field required to resolve the finite temperature-dependent width of these transitions. Figure~\ref{fig: D2sigmaTransitionsCrossing}a shows the second cell absorption profile, using the parameters of the Voigt cell from figure~\ref{fig: RbLineCentre}. The corresponding optical depth $\alpha L$ is plotted in figure~\ref{fig: D2sigmaTransitionsCrossing}b, where the relationship between these subplots is transmission $\mathcal{T}$~=~\rm{exp}$(-\alpha L)$~\cite{pizzey2022laser}. The total profile (grey) is calculated by summing the contributions from each transition; see equation~\ref{eqn:alpha} ($^{85}$Rb blue, $^{87}$Rb red. Only $^{87}$Rb is shown in figure~\ref{fig: D2sigmaTransitionsCrossing}a for clarity). It is clear that the transmission window in the Rb spectrum is caused by resolving the two $^{87}$Rb transition profiles closest to line centre, which have been highlighted in the figure (red). We model these as Gaussian profiles, and a simple approximation can be used to determine the width where the profiles overlap. To determine the FWHM of a Gaussian profile, we set the exponential in equation~\ref{eqn:Gaussian} equal to 0.5. Equivalently, our problem must solve
\begin{equation} \label{eqn: transmissionWindowAlpha}
\frac{2\,(\alpha L)_{\rm{min}}}{(\alpha L)_{\rm{max}}} = \frac{-\rm{ln}(\mathcal{T}_{\rm{TW}})}{-\rm{ln}(\mathcal{T}_{\rm{min}})} = \rm{exp}[-(\Delta/\omega_{D})^{2}4\ln{2}]~,
\end{equation}
where $(\alpha L)_{\rm{min}}$ is the optical depth of each resonance at the point the two profiles overlap, $\Delta = 0$ at $(\alpha L)_{\rm{max}}$, and $\mathcal{T}_{\rm{TW}}$ is used to denote the transmission at the peak of the transmission window. Equation~\ref{eqn: transmissionWindowAlpha} assumes two symmetrically spaced profiles of equal height, which are both true at sufficiently large $B$. The width at the overlap is therefore
\begin{equation}
\Delta = \frac{\omega_{\rm{D}}}{2}\sqrt{\frac{-1}{\rm{ln}(2)}\rm{ln}\Bigg[\frac{\rm{ln}(\mathcal{T}_{\rm_{TW}})/2}{\rm{ln}(\mathcal{T}_{\rm_{min}})}\Bigg]} > 1.15~\omega_{\rm{D}}~,
\end{equation}
where we have used $\mathcal{T}_{\rm{min}} < 0.001$ and $\mathcal{T}_{\rm{TW}} > 0.7$ for both transition profiles by observing figure~\ref{fig: D2sigmaTransitionsCrossing}a. Using the same $\mathcal{T}_{\rm{min}}$, $\Delta < 1.5~\omega_{\rm{D}}$ for $\mathcal{T}_{\rm{TW}} < 0.97$, and therefore this model satisfies the Gaussian approximation in Section~\ref{Sec: Theory}. In accordance with equation~\ref{eqn: strongTransitionEnergy}, $\Delta \sim (1/3)\mu_{\rm{B}}B$ for D2-$\pi$ transitions. We therefore approximate the thermal contribution $B_{\rm{th}}$ to $B(T)$ as
\begin{equation}
B_{\rm{th, D2}}^{\pi} > 2.46~\omega_{\rm{D}}~.
\end{equation}
This value is a minimum, since we use upper and lower limits of $\mathcal{T}_{\rm{min}}$ and $\mathcal{T}_{\rm{TW}}$ respectively when calculating $\Delta$. The simplicity in this model is that it is solely based on resolving optically thick transitions in the HPB regime, and applies to all alkali metals in this paper. By fixing $\mathcal{T}_{\rm{min}}$ and varying $\Delta$ in equation~\ref{eqn: transmissionWindowAlpha}, we expect $\mathcal{T}_{\rm{TW}}$ to trace out a Gaussian behaviour. The main temperature contribution is embedded in $\mathcal{T}_{\rm{min}}$, which is a simple approximation to $(\alpha L)_{\rm{max}}$ without the need to explicitly calculate it using equation~\ref{eqn:alpha}. A smaller $\mathcal{T}_{\rm{min}}$, and therefore larger $(\alpha L)_{\rm{max}}$ results in a larger width at a fixed $(\alpha L)_{\rm{min}}$. The consequence is a larger magnetic field to resolve the profiles. By summing $B_{\rm{SC}}$ and $B_{\rm{th}}$, we find a simple approximation to the second cell magnetic field of our cascaded filters for each alkali metal, transition and D line. This will be used to reduce two-cell parameter space in Section~\ref{Sec: OptimizedFilters}.

\subsection{Transmission window evolution}
\begin{figure}
\centering
{\includegraphics[width=\linewidth]{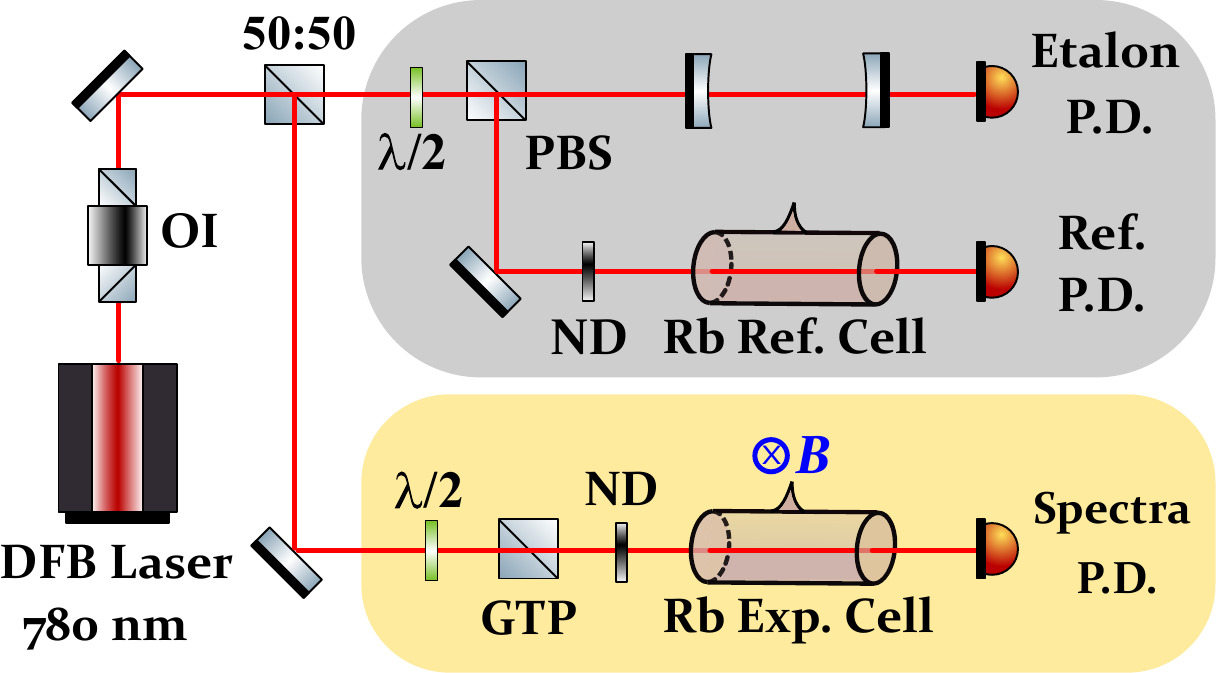}}
\caption{A schematic of the transmission window experiment. A distributed feedback (DFB) diode laser emits light resonant with the Rb-D2 line. After traversing an optical isolator (OI), light is split by a 50:50 beam splitter cube between reference optics (grey) and experiment optics (yellow). Before the $2~\rm{mm}$ experimental cell, horizontally polarised light is produced by a Glan-Taylor polariser. Light then traverses the cell, which is placed in a copper heater and situated within a transverse magnetic field (Voigt). The signal is then detected by a photodetector (P.D). Neutral density filters (ND) and half waveplates ($\lambda$/2) are used to control laser beam power. We use powers of order 100~\rm{nW}, ensuring an atom-light system that is in the weak probe regime~\cite{sherlock2009weak}. Reference optics are used as an atomic reference and to correct the non-linear laser scan~\cite{pizzey2022laser}. PBS: polarising beam splitter.}
\label{fig: experimentSetup}
\end{figure}

\begin{figure*}[t]
\centering
{\includegraphics[width=1.0\linewidth]{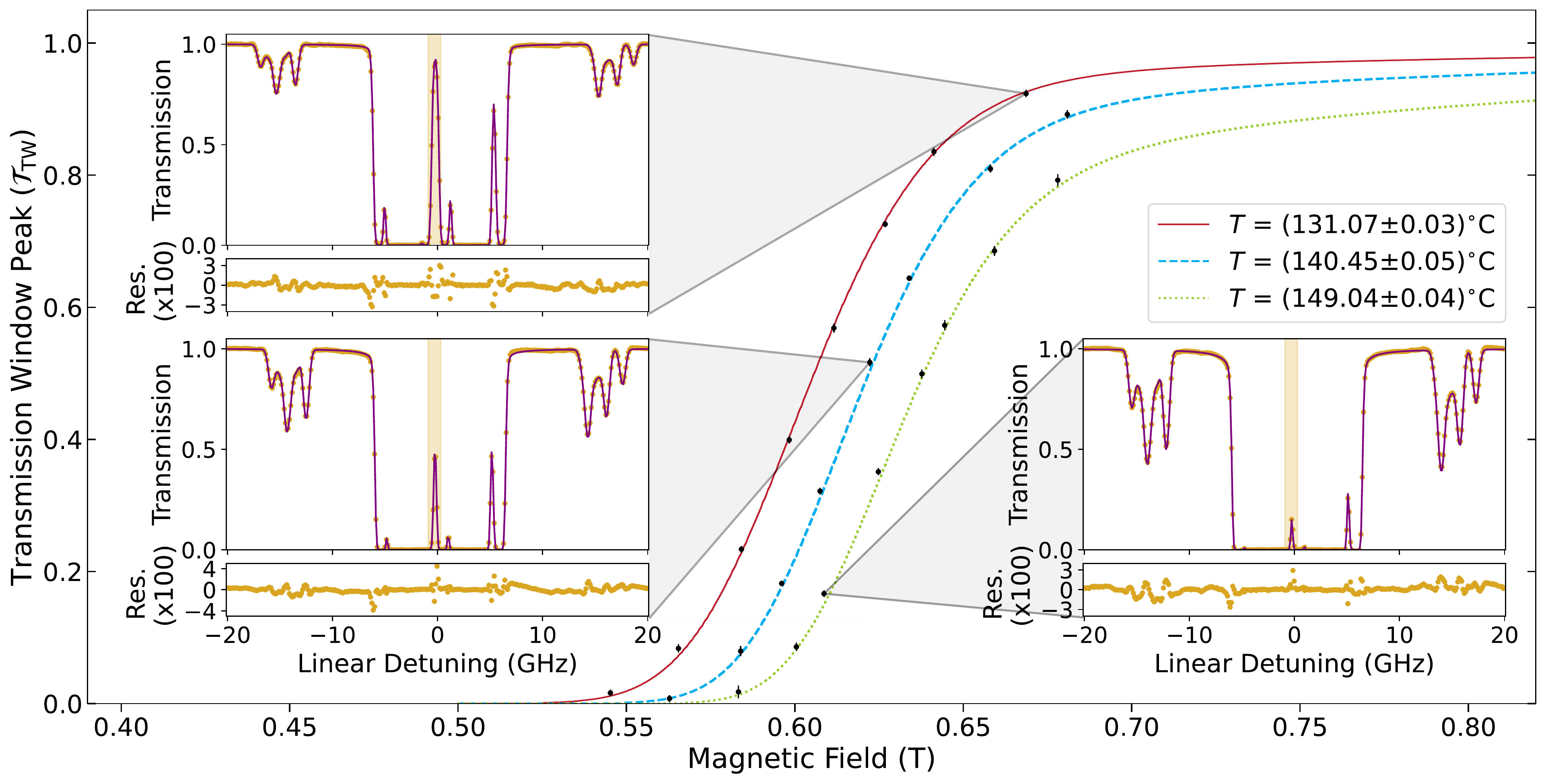}}
\caption{Evolution of line centre transmission as a function of magnetic field, thus demonstrating a transmission window $\mathcal{T}_{\rm{TW}}$ opening. Light traverses an optically thick 2~\rm{mm} natural abundance Rb vapour cell in the Voigt geometry ($\vec{k}\perp\vec{B}$). Each spectrum exhibits thermally broadened $\pi$-transition profiles which resolve at large magnetic fields. Three data sets are taken at constant temperatures $T = (131.07\pm0.03)^{\circ}\rm{C}$, $(140.45\pm0.05)^{\circ}\rm{C}$ and $(149.04\pm0.04)^{\circ}\rm{C}$, while increasing the magnetic field $B$. The insets show experiment (gold) vs theory (purple) for three spectra, while highlighting the line centre transmission region (yellow). $ElecSus$ interfaced with Marquardt-Levenberg fitting yields the best fit parameters~\cite{keaveney2018elecsus, zentile2015elecsus}, and from the fit we extract $T$ and $B$, while the data gives $\mathcal{T}_{\rm{TW}}$. The standard error in $B$ and $\mathcal{T}_{\rm{TW}}$ are shown as error bars and correspond to 4 repeats at each point. The error in $T$ corresponds to the standard error across a full data set, and is represented by broadened theory curves calculated using $ElecSus$. Both spectra residuals and line centre transmission data show excellent agreement between theory and experiment. }
\label{fig: experiment}
\end{figure*}

An experiment was set up to display the evolution of a Voigt transmission window as a function of magnetic field. The schematic is shown in figure~\ref{fig: experimentSetup}. A laser beam resonant with the Rb-D2 line is split between experimental and reference optics by a 50:50 beam splitter cube. Along the experimental channel, light is directed through a Glan-Taylor polariser, with optical power controlled by a half waveplate and neutral density filters. Horizontally polarised light then passes through a $2~\rm{mm}$ Rb vapour cell of natural abundance, situated within a resistive heater. A magnetic field is produced by two permanent NdFeB top hat shaped magnets, positioned either side of the cell and aligned in the Voigt geometry. The atom-light configuration is chosen such that $\pi$ transitions are induced. We control temperature by adjusting the current into the heater, and magnetic field via magnet separation. A $2~\rm{mm}$ experimental cell is used due to its narrower spatial extent. We therefore achieve an upper magnetic field limit of $\sim 0.68~\rm{T}$ while maintaining field homogeneity at the $<1\%$ level~\cite{keaveney2018optimized}. At $0.68~\rm{T}$, each Rb spectra typically covers $\sim 40~\rm{GHz}$. This range is achieved using a distributed feedback (DFB) laser whose central wavelength of $780~\rm{nm}$ is tuned via a temperature controller and has been shown to achieve a mode-hop free range of over $100~\rm{GHz}$~\cite{ponciano2020absorption}. The optical powers through experimental and reference cells were $700~\rm{nW}$ and $100~\rm{nW}$ respectively, with corresponding 1/$\rm{e}^{2}$ beam waists of $(665~\pm~3)~\rm{\upmu m}$ and $(846~\pm~4)~\rm{\upmu m}$. This ensures spectroscopy in the weak probe regime~\cite{sherlock2009weak}.

Each Rb spectrum was linearised using the combination of a Fabry-Pérot etalon and a $75~\rm{mm}$ natural abundance Rb reference cell, which has known features at room temperature and zero magnetic field. Similar analysis using a sub-Doppler atomic reference is described in~\cite{pizzey2022laser}. The spectra were fit using $ElecSus$ in order to extract the temperature and magnetic field of the atoms, which are initially treated as floating parameters. A total of three data sets were taken, each at different constant $T$ (fix current into heater) while varying $B$ (magnet separation). For each fixed $B$, four spectra were taken to account for random errors~\cite{hughes2010measurements}.  In order to account for small temperature variations, we calculate $T_{\rm{m}}$ as the mean temperature of all fits in a data set, along with its corresponding standard error~\cite{hughes2010measurements}. Each spectrum is then refitted using $T_{\rm{m}}$ to determine the best fit magnetic field of each spectrum. The mean magnetic field $B_{\rm{m}}$ and its error are then taken across each set of four repeats. To calculate the peak of the transmission window, which we denoted $\mathcal{T}_{\rm{TW}}$ in Section~\ref{Subsec: Model}, we first calculate the expected detuning of $\mathcal{T}_{\rm{TW}}$ using $T_{\rm{m}}$, $B_{\rm{m}}$ and $ElecSus$. Each normalised spectrum has 100,000 data points, from which we take the mean transmission in the $5~\rm{MHz}$ vicinity of this expected detuning to determine $\mathcal{T}_{\rm{TW}}$. We again take mean values and errors across the four repeats to determine $\mathcal{T}_{\rm{TW}}$ at $T_{\rm{m}}$ and $B_{\rm{m}}$.

The data are plotted in figure~\ref{fig: experiment}, along with theory curves showing the expected evolution of $\mathcal{T}_{\rm{TW}}$ vs $B$. These were calculated using $ElecSus$. The simple two transition model is visible from the figure: an atomic contribution where two profiles cross, causing a minimum $B$ which acts to shift the sigmoidal curves horizontally, and a second thermal contribution whose evolution depends on both atomic Doppler widths and transition line gradients. Similar sigmoidal behaviour was shown in~\cite{kiefer2014faraday}.
At lower temperatures, we see the evolution is near Gaussian. As discussed in the previous section, increasing temperature has a concomitant increase in number density and $(\alpha L)_{\rm{max}}$, which also acts to shift the curves horizontally to larger $B$. The effect of additional Lorentzian broadening at larger temperatures manifests itself in increased off-resonance absorption~\cite{weller2011absolute}. Consequently, the behaviour of the curves beyond $\mathcal{T}_{\rm{TW}} > 0.7$ more closely follows $1/\Delta^{2}$~\cite{siddons2009off} than the Gaussian behaviour $\rm{exp}(-\Delta^{2})$ in equation~\ref{eqn:Gaussian}. The large optical depth also means that, even off-resonance, there are fractional contributions from other resonances (see inset of figure ~\ref{fig: D2sigmaTransitionsCrossing}b). Further deviations from the model come from the predicted transition crossing in the HPB regime. We therefore expect that our model produces similar curves to those in figure~\ref{fig: experiment} but with steeper gradients, and shifted in magnetic field. It is clear from figure~\ref{fig: experiment} that an optically thick thermal vapour in the Voigt geometry creates a narrow transmission window when a sufficiently large magnetic field is applied.

\section{Optimised line centre filters}
\label{Sec: OptimizedFilters}

\begin{figure*}[t]
\centering
{\includegraphics[width=\linewidth]{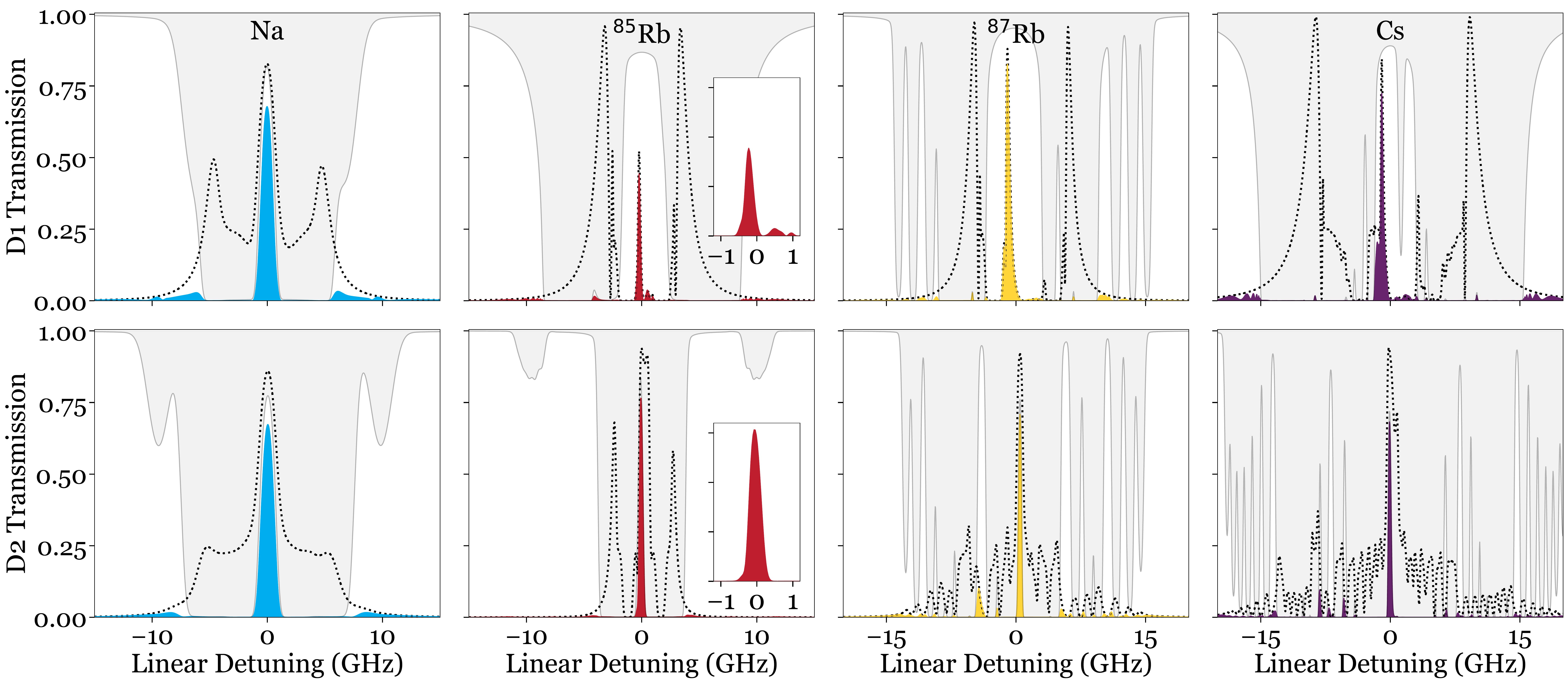}}
\caption{Optimised cascaded line centre filters in the Faraday-Voigt cell configuration. In colour, we show output filter profiles for Na, $^{85}$Rb, $^{87}$Rb and Cs (columns) for both D lines (rows). The second cell (Voigt) spectrum is plotted (grey), which acts as an optically thick window for first cell transmission (Faraday filter without second cell, black dotted). Transitions across the second cell are related to the angle of input polarisation~\cite{ponciano2020absorption}. The $^{85}$Rb second cell predominantly exhibits $\pi$ transitions, whereas the others predominantly exhibit $\sigma^{+/-}$ transitions. Parameters of each filter are shown in Table~\ref{Table: filters}.}
\label{fig: CascadedFilter}
\end{figure*}

Our theoretical model was used to investigate cascaded cell atomic filters in the Faraday-Voigt configuration for Na, K, Rb, and Cs across both D lines. This model has been experimentally verified~\cite{logue2022better}. Magnetic field angle $\theta_{B}$ is fixed by the cell configuration (see figure~\ref{fig: RbLineCentre}). We also fix element, isotopic abundance, D line, and cell length. Cell lengths of $5~\rm{mm}$ were chosen as a compromise between optical depth and magnetic field homogeneity, although recent literature has shown it is possible to attain tuneable homogeneous kilogauss magnetic fields for cell lengths of 25~\rm{mm}~\cite{pizzey2021tunable}. Each cascaded filter has a FOM which varies non-trivially across a multi-dimensional parameter space spanning both cells. We therefore interface $ElecSus$ with an optimisation program based on the SciPy differential evolution algorithm to find parameters which maximise the FOM of each filter. We calculate FOM using a linear detuning grid which spans $\pm50~\rm{GHz}$ in $10~\rm{MHz}$ steps, with an additional 5000 points centered in a $\pm0.5~\rm{GHz}$ range around each filter's peak transmission. The optimisation routine returns local maxima, and is therefore iterated to determine the cell parameters which give a global FOM maximum. We allow 5 parameters to vary: these are $T_{1}$, $B_{1}$, $T_{2}$, $B_{2}$ and $\theta_{E}$. Filters with either $\sigma^{+/-}$  or $\pi$ transitions across the second cell were investigated, and therefore $\theta_{E}$ was constrained in a $\pm 5^{\circ}$ range about $0^{\circ}$ or $90^{\circ}$ respectively. This also satisfies the requirements for an optically thick medium (see Section~\ref{Sec: TwoCells}). By constraining $\theta_{E}$, we are also able to add a minimum constraint to $B_{2}$ using the model in Section~\ref{Subsec: Model}, which requires knowledge of the second cell transitions to calculate $B_{\rm{SC}}$. By assuming Gaussian line shapes, room temperature Doppler widths $\omega_{\rm{D}}$, and applying a 1/1.217 correction to the strong cross component, we expect the model to underestimate $B_{2}$ required to open the transmission window. The constraints typically yielded a $\sim 30\%$ reduction in optimisation time for any filter run\footnote{Using a computer with an Intel Core i5-1135G7 processor. An unconstrained run would typically take $80-100$ minutes, and for each filter we execute a minimum of 10 runs.}. A summary of the optimised parameters can be found in Table~\ref{Table: filters}, with a selection of filters shown in figure~\ref{fig: CascadedFilter}. There was no obvious connection as to whether Faraday-Voigt line centre filters were better with $\pi$ or $\sigma^{+/-}$ transitions across their second cells, so the table only shows filters with the best FOMs.

We find that Voigt transmission windows in optically thick atomic vapours provide great utility in eliminating the multi-peak structure of single cell atomic filters~\cite{gerhardt2018anomalous, zentile2015atomic} while maintaining ultra-narrow bandwidths and high peak transmissions. In most cases, $>80\%$ of the total transmission can be found in a single line centre peak whose FWHM is much less than a Doppler width. All cascaded line centre filters in Table~\ref{Table: filters} are better than the best single cell Faraday filters in current literature~\cite{zentile2015atomic}, with notable FOM improvements by factors of $4.7$ and $4.3$ for the natural abundance Rb and $^{85}$Rb D2 filters respectively. With the exception of Cs, these filters also improve upon single cell filters in unconstrained $\theta_{B}$ geometries, in most cases by a factor of 2~\cite{keaveney2018optimized}. The unconstrained geometry utilises unique atom-light interactions to narrow the width of filter profiles~\cite{future}, but this only applies to Rb and Cs~\cite{keaveney2018optimized}. For Na and K, atomic filters are limited by large Doppler widths. Therefore, an optically thick transmission window provides a unique method to create an ultra-narrow line centre filter. A key observation from figure~\ref{fig: CascadedFilter} is the symmetry of each second cell absorption profile, which we expect due to the symmetric nature of alkali metal transitions in the HPB regime. The main differentiating factor is that $^{87}$Rb and Cs are more difficult to make optically thick due to their small Doppler widths. It has been shown that larger optical powers reduce optical depths as less atoms are in the ground state absorbing light~\cite{Siddons_2008}. This can be compensated for by larger number densities $\mathcal{N}(T)$. Assuming the same symmetric transition structure, we therefore expect optically thick transmission windows to have future utility in significantly reducing the influence of signal intensity on line centre filters beyond the weak probe regime~\cite{luo2018signal}.

\begin{table*}
\caption{\label{Table: filters}Optimised line centre filter parameters for light cascaded through both a Faraday ($L_{1} = 5~\rm{mm}$) and Voigt ($L_{2} = 5~\rm{mm}$) thermal vapour cell. The percentage of total filter transmission $\mathcal{T}$ contained within a single peak at line centre is quantified by $\mathcal{T}_{\rm{SP}}$. FWHMs are in units of Doppler widths, where $\omega_{\rm{D}}$ is calculated using first cell temperature $T_{1}$.}
\begin{indented}
\item[]\begin{tabular}{@{}lllllllllll}
\br
Elem & D line & $T_{1}$ ($^{\circ}\rm{C}$) & $B_{1}$ ($\rm{mT}$) & $T_{2}$ ($^{\circ}\rm{C}$) & $B_{2}$ ($\rm{mT}$) & $\theta_{E}$ ($\rm{^{\circ}}$) & FOM ($\rm{GHz}^{-1}$) & FWHM $(\omega_{\rm{D}}$) & $\mathcal{T}_{\rm{max}}$ & $\mathcal{T}_{\rm{SP}}$ (\%)\\ 

\mr
K$^{\rm{a}}$ & D2 & $\lineup$ 109.8 & $\lineup\0$ 82.9 & $\lineup$ 155.0 & $\lineup$ 110.4 & $\lineup\0$ 3.1 & $\lineup$ 1.05 & $\lineup$ 0.64 & $\lineup$ 0.69 & $\lineup$ 87.0\\
Na & D2 & $\lineup$ 193.7 & $\lineup$ 176.7 & $\lineup$ 244.2 & $\lineup$ 229.1 & $\lineup\0$ 3.3 & $\lineup$ 0.48 & $\lineup$ 0.73 & $\lineup$ 0.67 & $\lineup$ 85.4\\
Rb$^{\rm{a}}$ & D2 & $\lineup$ 105.9 & $\lineup\0\0$ 8.4 & $\lineup$ 128.7 & $\lineup$ 658.4 & $\lineup$ 90.6 & $\lineup$ 1.88 & $\lineup$ 0.57 & $\lineup$ 0.71 & $\lineup$ 92.0\\
$^{85}$Rb & D2 & $\lineup\0$ 95.3 & $\lineup\0$ 14.3 & $\lineup$ 105.8 & $\lineup$ 419.6 & $\lineup$ 92.3 & $\lineup$ 1.92 & $\lineup$ 0.62 & $\lineup$ 0.77 & $\lineup$ 91.8\\
$^{87}$Rb & D2 & $\lineup\0$ 82.8 & $\lineup$ 253.0 & $\lineup$ 123.1 & $\lineup$ 261.9 & $\lineup\0$ 1.0 & $\lineup$ 1.09 & $\lineup$ 0.70 & $\lineup$ 0.71 & $\lineup$ 60.4\\
Cs & D2 & $\lineup\0$ 72.6 & $\lineup$ 357.3 & $\lineup$ 120.1 & $\lineup$ 362.3 & $\lineup\0$ 1.5 & $\lineup$ 1.10 & $\lineup$ 0.88 & $\lineup$ 0.69 & $\lineup$ 58.9\\ 
\mr
K$^{\rm{a}}$ & D1 & $\lineup$ 120.7 & $\lineup\0$ 64.0 & $\lineup$ 164.3 & $\lineup$ 165.7 & $\lineup$ 94.5 & $\lineup$ 1.03 & $\lineup$ 0.61 & $\lineup$ 0.70 & $\lineup$ 81.4\\
Na & D1 & $\lineup$ 209.8 & $\lineup$ 139.2 & $\lineup$ 271.0 & $\lineup$ 349.0 & $\lineup$ 93.7 & $\lineup$ 0.49 & $\lineup$ 0.65 & $\lineup$ 0.68 & $\lineup$ 79.3\\
Rb$^{\rm{a}}$ & D1 & $\lineup$ 131.9 & $\lineup\0$ 19.8 & $\lineup$ 155.5 & $\lineup$ 563.7 & $\lineup$ 91.2 & $\lineup$ 1.17 & $\lineup$ 0.44 & $\lineup$ 0.48 & $\lineup$ 69.7\\
$^{85}$Rb & D1 & $\lineup$ 128.0 & $\lineup\0$ 20.0 & $\lineup$ 171.0 & $\lineup$ 336.4 & $\lineup\0$ 1.1 & $\lineup$ 1.04 & $\lineup$ 0.42 & $\lineup$ 0.44 & $\lineup$ 64.5\\
$^{87}$Rb & D1 & $\lineup$ 124.8 & $\lineup\0$ 31.6 & $\lineup$ 156.6 & $\lineup$ 645.3 & $\lineup$ 90.9 & $\lineup$ 1.15 & $\lineup$ 0.84 & $\lineup$ 0.83 & $\lineup$ 86.7\\
Cs & D1 & $\lineup$ 102.6 & $\lineup$ 150.5 & $\lineup$ 158.1 & $\lineup$ 527.3 & $\lineup\0$ 2.4 & $\lineup$ 0.86 & $\lineup$ 0.83 & $\lineup$ 0.73 & $\lineup$ 65.1\\ 
\br
\end{tabular}
\item[] $^{\rm{a}}$ Natural abundance isotopic ratio.
\end{indented}
\end{table*}

\section{Conclusion}
In this work we showed ultra-narrow transmission windows can be created in optically thick atomic vapours via spectroscopy in the Voigt geometry. The direct application is a second cell in a cascaded cell atomic filter, whose role is to remove noise displaced away from line centre. We showed the window opening mechanism, by applying a large magnetic field to resolve the two thermally broadened transitions closest to zero detuning. In this regime, a simple two transition model was calculated to describe the transmission window evolution. A near Gaussian evolution was demonstrated via experiment, and deviations from the model were discussed. Finally, we showed theoretical optimised cascaded filters for Na, K, Rb and Cs across both D lines, which form in transmission windows and output the majority of light through a single peak at line centre. These filters have large peak transmissions and bandwidths much less than a Doppler width.

\ack
The authors acknowledge EPSRC for funding this work (Grant No. EP/R002061/1). 

\section*{Disclosures}
The authors declare no conflicts of interest.

\section*{Data availability}
Data underlying the results presented in this paper are available from~\cite{data}.

\appendix
\section{Crossing of Transition Frequencies in the HPB Regime} \label{App1}

In this appendix, we derive equation~\ref{eqn: strongPiCrossD2}, the magnetic field where the two D2-$\pi$ transition energies closest to zero detuning cross in the HPB regime. In Section~\ref{Sec: Theory}, we wrote the eigenenergy approximation of an eigenstate $\vert I, m_{I}, J, m_{J} \rangle$ in the HPB regime
\begin{eqnarray}
E_{\vert I, m_{I}, J, m_{J} \rangle} = A\,m_{I}m_{J} + \mu_{\rm{B}}B(g_{I}m_{I} + g_{J}m_{J}).
\label{APDXeqn: strongEnergy1}
\end{eqnarray}
Omitting the uncoupled basis state notation, we start by using equation~(\ref{APDXeqn: strongEnergy1}) to derive the energy of an allowed electric-dipole transition
\begin{eqnarray}
\Delta E & = m_{I}(A'\,m'_{J} - A\,m_{J}) \\ \nonumber 
& + \mu_{\rm{B}}B(g'_{J}m'_{J} - g_{J}m_{J})~,
\label{APDXeqn: strongEnergy2}
\end{eqnarray}
where primes indicate the excited state, $g'_{I} = g_{I}$, and the selection rule $\Delta m_{I} = 0$ has been used~\cite{woodgate1970elementary}. Next we make the substitution $m'_{J} = m_{J} + q$:
\begin{eqnarray}\label{APDXeqn: strongEnergy3}
\Delta E(q) & = m_{I}[m_{J}(A' - A) + qA'] \\ \nonumber  
& + \mu_{\rm{B}}B[m_{J}(g'_{J} - g_{J}) + qg'_{J}]~.
\end{eqnarray}
By using this notation, we have $\sigma^{\pm}$ transitions when $q\equiv \Delta m_J = m'_{J}-m_{J} = \pm1$,  and $\pi$ transitions when $\Delta m_{J} = 0$. To determine the magnetic field at transition energy crossings, we need to solve $\Delta E_{1}(q_{1}) = \Delta E_{2}(q_{2})$. The solutions correspond to the transitions $\vert I, m_{I_{1}}, 1/2, m_{J_{1}} \rangle \rightarrow \vert I, m_{I_{1}}, J', m_{J_{1}}+q_{1} \rangle$ and $\vert I, m_{I_{2}}, 1/2, m_{J_{2}} \rangle \rightarrow \vert I, m_{I_{2}}, J', m_{J_{2}}+q_{2} \rangle$ respectively:
\begin{eqnarray}\label{APDXeqn: BigBeqn}
B = \frac{\Delta E_{1, \rm{hfs}}(q_{1}) - \Delta E_{2, \rm{hfs}}(q_{2})}{\mu_{\rm{B}}[(g'_{J} - g_{J})(m_{J_{2}} - m_{J_{1}}) + g'_{J}(q_{2} - q_{1})]}~, \\ \nonumber
\Delta E_{1, \rm{hfs}}(q_{1}) = m_{I_{1}}[m_{J_{1}}(A' - A) + q_{1}A']~, \\ \nonumber
\Delta E_{2, \rm{hfs}}(q_{2}) = m_{I_{2}}[m_{J_{2}}(A' - A) + q_{2}A']~.
\end{eqnarray}
The case we are considering is the splitting of groups of $\pi$ transitions on the D2 line. In the HPB regime, there are two groups of $\pi$ transitions between the ground and excited state manifolds
\begin{eqnarray}
\vert I, m_{I_{1}}, 1/2, 1/2 \rangle &\rightarrow & \vert I, m_{I_{1}}, 3/2, 1/2 \rangle~, \\ \nonumber
\vert I, m_{I_{2}}, 1/2, -1/2 \rangle &\rightarrow & \vert I, m_{I_{2}}, 3/2, -1/2 \rangle ~.
\label{APDXeqn: strongSigmaTransitions}
\end{eqnarray}
By substituting $q_{1} = q_{2} = 0$, $g'_{J} \sim 4/3$ and $g_{J} \sim 2$~\cite{etheses7747} into equation~\ref{APDXeqn: strongEnergy3}, we find the two transition groups have gradients $\pm (1/3)\mu_{\rm{B}}$ with respect to $B$:
\begin{eqnarray}\label{APDXeqn: strongPiD2TransitionEnergies}
\Delta E_{1}(q_{1} = 0) &=& \frac{m_{I_{1}}}{2}(A' - A) - \frac{1}{3}\mu_{\rm{B}}B~, \\ \nonumber
\Delta E_{2}(q_{2} = 0) &=& -\frac{m_{I_{2}}}{2}(A' - A) + \frac{1}{3}\mu_{\rm{B}}B~.
\end{eqnarray}
The first terms in equations~\ref{APDXeqn: strongEnergy3} and \ref{APDXeqn: strongPiD2TransitionEnergies} are related to the hyperfine interaction, and correspond to the splitting within each transition group. We require the strong transition within each group closest to zero detuning as these are responsible for opening a transmission window. Inspection of the aforementioned equations gives $m_{I_{1}} = m_{I_{2}} = -I$, and therefore we solve equations~\ref{APDXeqn: strongPiD2TransitionEnergies} simultaneously:
\begin{equation}
B_{\rm{SC, D2}}^{\rm{\pi}} = \frac{3}{2}\frac{(A - A')I}{\mu_{\rm{B}}}~,
\label{APDXeqn: strongPiCrossD2}
\end{equation}
where SC is strong cross, and D2 corresponds to the D line of the $\pi$ transitions. The following equations can be derived in the same way:
\begin{eqnarray}
B_{\rm{SC, D2}}^{\sigma} = \frac{1}{2}\frac{(A - 3A')I}{\mu_{\rm{B}}}~, \\
B_{\rm{SC, D1}}^{\pi} = \frac{3}{4}\frac{(A - A')I}{\mu_{\rm{B}}}~, \\
B_{\rm{SC, D1}}^{\sigma} = \frac{3}{8}\frac{(A + A')I}{\mu_{\rm{B}}}~. 
\label{APDXeqn: strongCrossOther}
\end{eqnarray}
By identifying the quantum numbers responsible for the transitions closest to zero detuning, equations~\ref{APDXeqn: strongPiCrossD2}-~\ref{APDXeqn: strongCrossOther} can also be found by direct substitution into equation~\ref{APDXeqn: BigBeqn}.

\section*{References}

\bibliographystyle{unsrt}
\bibliography{bibliography.bib}

\begin{thebibliography}{10}

\bibitem{caltzidis2021atomic}
Caltzidis I, K{\"u}bler H, Pfau T, L{\"o}w R, and Zentile M~A.
\newblock {\em Phys. Rev. A}, 103(4):043501, 2021.

\bibitem{higgins2021electromagnetically}
Higgins C~R and Hughes I~G.
\newblock {\em J. Phys. B: At. Mol. Opt. Phys.}, 54(16):165403, 2021.

\bibitem{siverns2019demonstration}
Siverns J~D, Hannegan J, and Quraishi Q.
\newblock {\em Sci. Advances}, 5(10):eaav4651, 2019.

\bibitem{budker2002resonant}
Budker D, Gawlik W, Kimball D~F, Rochester S~M, Yashchuk V~V, and Weis A.
\newblock {\em Rev. Mod. Phys}, 74(4):1153, 2002.

\bibitem{auzinsh2010optically}
Auzinsh M, Budker D, and Rochester S.
\newblock {\em Optically Polarized Atoms: Understanding Light-Atom
  Interactions}.
\newblock Oxford University Press, 2010.

\bibitem{weller2012optical}
Weller L, Kleinbach K~S, Zentile M~A, Knappe S, Hughes I~G, and Adams C~S.
\newblock {\em Opt. Lett.}, 37(16):3405--3407, 2012.

\bibitem{aplet1964faraday}
Aplet L~J and Carson J~W.
\newblock {\em Appl. Opt.}, 3(4):544--545, 1964.

\bibitem{budker2007optical}
Budker D and M~Romalis.
\newblock {\em Nat. Phys.}, 3(4):227--234, 2007.

\bibitem{sutter2020recording}
Sutter J~U, Lewis O, Robinson C, et~al.
\newblock {\em Comput. Electron. Agric.}, 177:105651, 2020.

\bibitem{kitching2018chip}
Kitching J.
\newblock {\em Appl. Phys. Rev.}, 5(3):031302, 2018.

\bibitem{auzinsh2022wide}
Auzinsh M, Sargsyan A, Tonoyan A, Leroy C, Momier R, Sarkisyan D, and Papoyan
  A.
\newblock {\em Appl. Opt.}, 61(19):5749--5754, 2022.

\bibitem{nagourney2014quantum}
Nagourney W.
\newblock {\em Quantum Electronics for Atomic Physics and Telecommunication}.
\newblock OUP Oxford, 2014.

\bibitem{gerhardt2018anomalous}
Gerhardt I.
\newblock {\em Opt. Lett.}, 43(21):5295--5298, 2018.

\bibitem{dick1991ultrahigh}
Dick D~J and Shay T~M.
\newblock {\em Opt. Lett.}, 16(11):867--869, 1991.

\bibitem{menders1991ultranarrow}
Menders J, Benson K, Bloom S~H, Liu C~S, and Korevaar E.
\newblock {\em Opt. Lett.}, 16(11):846--848, 1991.

\bibitem{yin1991theoretical}
Yin B and Shay T~M.
\newblock {\em Opt. Lett.}, 16(20):1617--1619, 1991.

\bibitem{chen1993sodium}
Chen H, She C~Y, Searcy P, and Korevaar E.
\newblock {\em Opt. Lett.}, 18(12):1019--1021, 1993.

\bibitem{Harrell:09}
Harrell S~D, She C~Y, Yuan T, Krueger D~A, Chen H, Chen S~S, and Hu~Z~L.
\newblock {\em JOSA B}, 26(4):659--670, 2009.

\bibitem{yan2022dual}
Yan Y, Yuan J, Wang L, Xiao L, and Jia S.
\newblock {\em Opt. Commun.}, 509:127855, 2022.

\bibitem{menders1992blue}
Menders J, Searcy P, Roff K, and Korevaar E.
\newblock {\em Opt. Lett.}, 17(19):1388--1390, 1992.

\bibitem{kudenov2020dual}
Kudenov M~W, Pantalone B, and Yang R.
\newblock {\em Appl. Opt.}, 59(17):5282--5289, 2020.

\bibitem{yin2016tunable}
Yin L, Luo B, Xiong J, and Guo H.
\newblock {\em Opt. Express}, 24(6):6088--6093, 2016.

\bibitem{faraday1846experimental}
Faraday M.
\newblock {\em Phil. Trans. R. Soc. L.}, (136):1--20, 1846.

\bibitem{ponciano2020absorption}
Ponciano-Ojeda F~S, Logue F~D, and Hughes I~G.
\newblock {\em J. Phys. B: At. Mol. Opt. Phys.}, 54(1):015401, 2020.

\bibitem{keaveney2018self}
Keaveney J, Rimmer D~A, and Hughes I~G.
\newblock {\em arXiv preprint arXiv:1807.04652}, 2018.

\bibitem{higgins2020atomic}
Higgins C~R, Pizzey D, Mathew R~S, and Hughes I~G.
\newblock {\em OSA Contin.}, 3(4):961--970, 2020.

\bibitem{keaveney2018optimized}
Keaveney J, Wrathmall S~A, Adams C~S, and Hughes I~G.
\newblock {\em Opt. Lett.}, 43(17):4272--4275, 2018.

\bibitem{keaveney2018elecsus}
Keaveney J, Adams C~S, and Hughes I~G.
\newblock {\em Comput. Phys. Commun.}, 224:311--324, 2018.

\bibitem{rotondaro2015generalized}
Rotondaro M~D, Zhdanov B~V, and Knize R~J.
\newblock {\em JOSA B}, 32(12):2507--2513, 2015.

\bibitem{siyushev2014molecular}
Siyushev P, Stein G, Wrachtrup J, and Gerhardt I.
\newblock {\em Nature}, 509(7498):66--70, 2014.

\bibitem{portalupi2016simultaneous}
Portalupi S~L, Widmann M, Nawrath C, Jetter M, Michler P, Wrachtrup J, and
  Gerhardt I.
\newblock {\em Nat. Commun.}, 7:13632, 2016.

\bibitem{yong2011flat}
Yong Y, Xuewu C, Faquan L, Xiong H, Xin L, and Shunsheng G.
\newblock {\em Opt. Lett.}, 36(7):1302--1304, 2011.

\bibitem{li2012doppler}
Li~F~Q, Cheng X~W, Lin X, Yang Y, Wu~K~J, Liu. Y~J, Gong S~S, and Song S~L.
\newblock {\em Opt. \& Laser Tech.}, 44(6):1982--1986, 2012.

\bibitem{shi2020dual}
Shi T, Guan X, Chang P, Miao J, Pan D, Luo B, Guo H, and Chen J.
\newblock {\em IEEE Photonics Journal}, 12(4):1--11, 2020.

\bibitem{chang2022frequency}
Chang P, Shi H, Miao J, Shi T, Pan D, Luo B, Guo H, and Chen J.
\newblock {\em Appl. Phys. Lett.}, 120(14):141102, 2022.

\bibitem{tang202118w}
Tang H, Zhao H, Wang R, Li~L, Yang Z, Wang H, Yang W, Han K, and Xu~X.
\newblock {\em Opt. Express}, 29(23):38728--38736, 2021.

\bibitem{keaveney2016single}
Keaveney J, Hamlyn W~J, Adams C~S, and Hughes I~G.
\newblock {\em Rev. Sci. Instrum.}, 87(9):095111, 2016.

\bibitem{yin2022using}
Yin L, Qian D, Geng Z, Zhan H, and Wu~G.
\newblock {\em Opt. Express}, 30(20):36297--36306, 2022.

\bibitem{junxiong1995experimental}
Junxiong T, Qingji W, Yimin L, Liang Z, Jianhua G, Minghao D, Jiankun K, and
  Lemin Z.
\newblock {\em Appl. Opt.}, 34(15):2619--2622, 1995.

\bibitem{zhang2021background}
Zhang J, Gao G, Wang B, Guan X, Yin L, Chen J, and Luo B.
\newblock {\em J. Light. Technol.}, 40(1):63--73, 2021.

\bibitem{erdelyi2022solar}
Erd{\'e}lyi R et~al.
\newblock {\em J. Space Weather Space Clim.}, 12, 2022.

\bibitem{cacciani1978magneto}
Cacciani A and Fofi M.
\newblock {\em Sol. Phys.}, 59(1):179--189, 1978.

\bibitem{cacciani1990solar}
Cacciani A, Ricci D, Rosati P, Rhodes E~J, Smith E, Tomczyk S, and Ulrich R~K.
\newblock {\em Il Nuovo Cimento C}, 13(1):125--130, 1990.

\bibitem{logue2022better}
Logue F~D, Briscoe J~D, Pizzey D, Wrathmall S~A, and Hughes I~G.
\newblock {\em Opt. Lett.}, 47(12):2975--2978, 2022.

\bibitem{kiefer2014faraday}
Kiefer W, L{\"o}w R, Wrachtrup J, and Gerhardt I.
\newblock {\em Sci. Rep.}, 4:6552, 2014.

\bibitem{keaveney2019quantitative}
Keaveney J, Ponciano-Ojeda F~S, Rieche S~M, Raine M~J, Hampshire D~P, and
  Hughes I~G.
\newblock {\em J. Phys. B: At. Mol. Opt. Phys.}, 52(5):055003, 2019.

\bibitem{siddons2009off}
Siddons P, Adams C~S, and Hughes I~G.
\newblock {\em J. Phys. B: At. Mol. Opt. Phys.}, 42(17):175004, 2009.

\bibitem{weller2011absolute}
Weller L, Bettles R~J, Siddons P, Adams C~S, and Hughes I~G.
\newblock {\em J. Phys. B: At. Mol. Opt. Phys.}, 44(19):195006, 2011.

\bibitem{sargsyan2022saturated}
Sargsyan A, Momier R, Leroy C, and Sarkisyan D.
\newblock {\em Laser Phys.}, 32(10):105701, 2022.

\bibitem{sargsyan2012hyperfine}
Sargsyan A, Hakhumyan G, Leroy C, Pashayan-Leroy Y, Papoyan A, and Sarkisyan D.
\newblock {\em Opt. Lett.}, 37(8):1379--1381, 2012.

\bibitem{olsen2011optical}
Olsen B~A, Patton B, Jau Y-Y, and Happer W.
\newblock {\em Phys. Rev. A}, 84(6):063410, 2011.

\bibitem{zentile2014hyperfine}
Zentile M~A, Andrews R, Weller L, Knappe S, Adams C~S, and Hughes I~G.
\newblock {\em J. Phys. B: At. Mol. Opt. Phys.}, 47(7):075005, 2014.

\bibitem{weller2012absolute}
Weller L, Kleinbach K~S, Zentile M~A, Knappe S, Adams C~S, and Hughes I~G.
\newblock {\em J. Phys. B: At. Mol. Opt. Phys.}, 45(21):215005, 2012.

\bibitem{zentile2015atomic}
Zentile M~A, Whiting D~J, Keaveney J, Adams C~S, and Hughes I~G.
\newblock {\em Opt. Lett.}, 40(9):2000--2003, 2015.

\bibitem{guancold}
Guan X, Zhuang W, Shi T, Miao J, Zhang J, Chen J, and Luo B.
\newblock {\em Frontiers in Physics}, 10:1297, 2022.

\bibitem{zhuang2021ultranarrow}
Zhuang W, Zhao Y, Wang S, Fang Z, Fang F, and Li~T.
\newblock {\em Chin. Opt. Lett.}, 19(3):030201, 2021.

\bibitem{bi2016ultra}
Bi~G, Kang J, Fu~J, Ling L, and Chen J.
\newblock {\em Phys. Lett. A}, 380(47):4022--4026, 2016.

\bibitem{luo2018signal}
Luo B, Yin L, Xiong J, Chen J, and Guo H.
\newblock {\em Opt. Lett.}, 43(11):2458--2461, 2018.

\bibitem{zentile2015elecsus}
Zentile M~A, Keaveney J, Weller L, Whiting D~J, Adams C~S, and Hughes I~G.
\newblock {\em Comput. Phys. Commun.}, 189:162--174, 2015.

\bibitem{sherlock2009weak}
Sherlock B~E and Hughes I~G.
\newblock {\em Am. J. Phys.}, 77(2):111--115, 2009.

\bibitem{etheses7747}
Weller L.
\newblock PhD thesis, Durham University, 2013.

\bibitem{Siddons_2008}
Siddons P, Adams C~S, Ge~C, and Hughes I~G.
\newblock {\em J. Phys. B: At. Mol. Opt. Phys}, 41(15):155004, 2008.

\bibitem{wu1986optical}
Wu~Z, Kitano M, Happer W, Hou M, and Daniels J.
\newblock {\em Appl. Opt.}, 25(23):4483--4492, 1986.

\bibitem{preston1996doppler}
Preston D~W.
\newblock {\em Am. J. Phys.}, 64(11):1432--1436, 1996.

\bibitem{lewis1980collisional}
Lewis E~L.
\newblock {\em Phys. Rep.}, 58(1):1--71, 1980.

\bibitem{tudor1963new}
Tudor~Davies J and Vaughan J~M.
\newblock {\em Astrophys. J}, 137:1302, 1963.

\bibitem{foot2004atomic}
Foot C~J.
\newblock {\em Atomic physics}, volume~7.
\newblock OUP Oxford, 2004.

\bibitem{woodgate1970elementary}
Woodgate G~K.
\newblock {\em Elementary Atomic Structure}.
\newblock Oxford Science Publications. Clarendon Press, 1980.

\bibitem{dressler1996theory}
Dressler E~T, Laux A~E, and Billmers R~I.
\newblock {\em JOSA B}, 13(9):1849--1858, 1996.

\bibitem{tremblay1990absorption}
Tremblay P, Michaud A, Levesque M, Th{\'e}riault S, Breton M, Beaubien J, and
  Cyr N.
\newblock {\em Phys. Rev. A}, 42(5):2766, 1990.

\bibitem{umfer1992investigations}
Umfer C, Windholz L, and Musso M.
\newblock {\em Z. Phys. D: At. Mol. Clus.}, 25(1):23--29, 1992.

\bibitem{windholz1985zeeman}
Windholz L.
\newblock {\em Z. Phys. A: At. Nuc.}, 322(2):203--206, 1985.

\bibitem{windholz1988zeeman}
Windholz L and Musso M.
\newblock {\em Z. Phys. D: At. Mol. Clus.}, 8(3):239--249, 1988.

\bibitem{george2017pulsed}
George S, Bruyant N, B{\'e}ard J, Scotto S, Arimondo E, Battesti R, Ciampini D,
  and Rizzo C.
\newblock {\em Rev. Sci. Instrum.}, 88(7):073102, 2017.

\bibitem{ciampini2017optical}
Ciampini D, Battesti R, Rizzo C, and Arimondo E.
\newblock {\em Phys. Rev. A}, 96(5):052504, 2017.

\bibitem{pizzey2022laser}
Pizzey D, Briscoe J~D, Logue F~D, Ponciano-Ojeda F~S, Wrathmall S~A, and Hughes
  I~G.
\newblock {\em New J. Phys.}, 2022.

\bibitem{hughes2010measurements}
Hughes I~G and Hase T.
\newblock {\em Measurements and their uncertainties: a practical guide to
  modern error analysis}.
\newblock OUP Oxford, 2010.

\bibitem{pizzey2021tunable}
Pizzey D.
\newblock {\em Rev. Sci. Instrum.}, 92(12):123002, 2021.

\bibitem{future}
Logue F~D, Briscoe J~D, Pizzey D, Wrathmall S~A, and Hughes I~G.
\newblock (in preparation).

\bibitem{data}
Briscoe J~D.
\newblock {Voigt transmission windows in optically thick atomic vapours: a
  method to create single-peaked line centre filters [dataset]. Durham
  University Collections.}
\newblock \url{http://doi.org/10.15128/r1j67313811}, 2022.

\end{thebibliography}

\end{document}